\documentclass[12pt,preprint]{aastex}
\input epsf

\newcommand\bs{\hspace{-0.25cm}}
\newcommand\bss{\hspace{-0.07cm}}

\newcommand\B{{\bf B}}
\newcommand\E{{\bf E}}
\newcommand\J{{\bf J}}

\newcommand\kk{{\bf k}}

\newcommand\V{{\bf V}}
\newcommand\A{{\bf A}}

\newcommand\bnabla{\mbox{\boldmath $\nabla$}}

\newcommand\smf{\scriptscriptstyle\mathsf}

\begin{document}

\nonumber
\setcounter{equation}{0}

\title{A constrained-transport magnetohydrodynamics algorithm with
near-spectral resolution.}


\author{Jason L. Maron\altaffilmark{1}}
\author{Mordecai-Mark Mac Low\altaffilmark{2}}
\author{Jeffrey S. Oishi\altaffilmark{3}}
\affil{Department of Astrophysics, American Museum of Natural History,
79th Street at Central Park West, New York, NY 10024-5192}
\altaffiltext{1}{badger@amnh.org}
\altaffiltext{2}{mordecai@amnh.org}
\altaffiltext{3}{Also Department of Astronomy, University of Virginia,
  Charlottesville, VA; joishi@amnh.org}

\begin{abstract}

Numerical simulations including magnetic fields have become important
in many fields of astrophysics.  Evolution of magnetic fields by the
constrained transport algorithm preserves magnetic divergence to
machine precision, and thus represents one preferred method for the
inclusion of magnetic fields in simulations.  We show that constrained
transport can be implemented with volume-centered fields and
hyperresistivity on a high-order finite difference
stencil. Additionally, the finite-difference coefficients can be tuned
to enhance high-wavenumber resolution. Similar techniques can be used
for the interpolations required for dealiasing corrections at high
wavenumber.  Together, these measures yield an algorithm with a
wavenumber resolution that approaches the theoretical maximum achieved
by spectral algorithms.  Because this algorithm uses finite
differences instead of fast Fourier transforms, it runs faster and
isn't restricted to periodic boundary conditions.  Also, since the
finite differences are spatially local, this algorithm is easily
scalable to thousands of processors.  We demonstrate that, for
low-Mach-number turbulence, the results agree well with a high-order,
non-constrained-transport scheme with Poisson divergence cleaning.

\end{abstract}

\section{Introduction} \label{ct}

Many astrophysical flows involve dynamically significant magnetic fields, such
as molecular clouds, accretion disks, the galactic dynamo, jets, galaxy
clusters, stellar dynamos and coronae, the solar wind and the interstellar
medium. These problems tend to be three-dimensional, multiscale and turbulent,
so there is an ongoing interest in developing high-resolution and efficient
magnetohydrodynamics (MHD) algorithms for them. In this paper, we
outline an extension of the constrained transport algorithm (Evans \& Hawley
1988) to the combination of higher spatial order and zone-centered grids, and
with resolution-enhanced tuned derivatives.  We then describe how these
measures fit together to yield an algorithm that closely approaches the
theoretical maximum wavenumber resolution of spectral algorithms.

\subsection{Constrained transport}

The induction equation for a magnetic field $\B$ and a velocity field $\V$
in ideal MHD is
\begin{equation}
\partial_t \B = \bnabla \times (\V \times \B). \label{eqct}
\end{equation}
Analytically, this equation conserves magnetic divergence: $\partial_t
(\bnabla \cdot \B) = 0$.  However, this may or may not be the case for
a finite-difference treatment of this equation.  T\'{o}th (2000)
reviews the methods taken by various algorithms to treat the
divergence in MHD simulations.  A spectral code explicitly projects
the Fourier components so that $\hat{\B}(\kk) \cdot \kk = 0$. For a
finite difference code, the magnetic field can be evolved by a
constrained transport scheme that preserves the magnetic divergence to
machine precision (Evans \& Hawley 1988).  Alternatively, if the
discretization doesn't conserve magnetic divergence, the divergence
can be removed with measures such as periodic use of a Poisson solver
(Brackbill \& Barnes 1980), adding a divergence diffusion term
$\partial_t \B = \bnabla \bnabla \cdot \B$ to the magnetic evolution,
or following an artificial and independently evolving divergence field
(Dedner et al.\ 2002) to propagate divergence away from where it is
produced and then dissipate it. The Powell scheme (Powell et al.\
1999) adds a source term to advect divergence rather than let it grow
in place.  A finite difference code can also employ a vector potential
${\bf A}$ such that $\B = \bnabla \times {\bf A}$, in which case the
magnetic divergence is automatically zero.  This requires the use of a
higher-order advection algorithm to ensure accurate
second-derivatives, as is done in the Pencil code (Brandenburg \&
Dobler 2002).

We denote any finite difference scheme for MHD that explicitly conserves the
magnetic divergence to machine precision
as constrained transport (CT), and any scheme that does not as
unconstrained transport (UT). Several variations of CT
are possible.  If the electric field is differenced as a curl: $\partial_t \B =
- \bnabla \times \E,$ 
then the magnetic divergence is preserved to machine
precision for most grid types (see Appendix).  Evans \& Hawley (1988)
introduced CT for staggered grids and T\'{o}th (2000)
showed that it works for centered grids as well (see the Appendix
for explanation of centered and staggered grids).  Londrillo \& Del Zanna
(2000) further showed that high-order CT is possible on
staggered grids with a radius-two stencil. In this paper, we show that
volume-centered CT is possible on arbitrarily large stencil
sizes, with hyperresistivity, and that the resolution of this algorithm
at moderately high order approaches the theoretical maximum exhibited by a
spectral code.

In \S~\ref{algorithms}, we discuss the specifics of the algorithm, and
in \S~\ref{simulations}, we describe test simulations that demonstrate
the capabilities of 
CT with high-order spatial derivatives.

\section{Algorithms} \label{algorithms}

Our algorithm is based on a constrained transport scheme, plus
measures to enhance the resolution and maintain stability.  High
wavenumber resolution is achieved by a combination of high-order and
tuned finite differences plus hyperdiffusivity, and stability is
achieved by Runge-Kutta timestepping and hyperdiffusivity.

\subsection{Timestepping}

A high-order timestepping scheme for the
evolution equations is essential for the
stability of most algorithms. The time update for a variable $Q(t)$ is
$Q(\Delta t) = Q(0) + \Delta t Q^*$, where $Q^*$ represents some
estimate of $\int_0^{\Delta t} \partial_t Q(t) dt$. One example is the
second-order Runge-Kutta scheme, which estimates $Q(\Delta t/2)
\approx Q(0) + (\Delta t / 2) Q^\prime(0),$ and then identifies $Q^* =
Q^\prime(\Delta t/2)$. Another class of algorithms maintains the
conservation of mass and momentum by computing fluxes through zone
boundaries. A variety of techniques exist for time-extrapolating the
fluxes at $t=\Delta t/2$, such as piecewise parabolic advection
(Colella and Woodward 1984), Total Variation Diminishing (Harten
1983), Riemann solvers (Toro 1999), the method of characteristics
(Stone \& Norman 1992a, Hawley \& Stone 1995), and many more. MHD
poses a challenge to time extrapolation because there are seven or
eight wavemode characteristics, depending on the technique used for
treating magnetic divergence.  In particular, the well-known
method-of-characteristics algorithm interpolates along the Alfv\'en
characteristic while neglecting the fast and slow mode
characteristics.  For our simulations, we use Runge-Kutta for
time-extrapolation because it doesn't invoke any diffusive spatial
interpolations (\S~\ref{diffusivity}), and because it automatically
captures all three MHD wavemode types.  In our demonstration
implementation, we use second-order Runge-Kutta while the Pencil code
(Brandenburg \& Dobler 2002) uses third-order, although either order has
proven successful.

\subsection{Diffusivity} \label{diffusivity}

A common class of algorithms is based on momentum fluxes that are
time-extrapolated with upwind spatial interpolations.  The errors from
the interpolations required for these flux transport algorithms
produce an intrinsic diffusivity that can stabilize the evolution,
even in the absence of any explicit diffusive terms.
The nature and magnitude of the diffusivity has been characterized in
Zhong (1998) and Dobler et al (2006).

Runge-Kutta timestepping, on the other hand, has no spatial
interpolations, and thus no intrinsic diffusivity.
One then generally needs an explicit stabilizing diffusivity.
One has various options for the form of this diffusivity, with
Laplacian or hyper-Laplacian typically chosen.
These diffusivities have the benefit that their magnitude is easily
characterized, and the diffusive coefficient can be tuned to have the
minimum value necessary to preserve stability. Consider:
\begin{equation}
\partial_t \V =
  \nu_{\smf 2} \bnabla^2 \V
- \nu_{\smf 4} \bnabla^4 \V
+ \nu_{\smf 6} \bnabla^6 \V
- \nu_{\smf [4]} \bnabla^{[4]} \V.
\end{equation}
Let the Fourier components be
$\hat{\V}(\kk),$ where $\kk = \{k_x, k_y, k_z\}$ is
the wavenumber.
They evolve according to
\begin{equation}
\partial_t \hat{\V}(\kk) =
- \nu_{\smf 2} \kk^2 \hat{\V}(\kk)
- \nu_{\smf 4} \kk^4 \hat{\V}(\kk)
- \nu_{\smf 6} \kk^6 \hat{\V}(\kk).
- \nu_{\smf [4]} (k_x^4 + k_y^4 + k_z^4) \hat{\V}(\kk).
\end{equation}
The $\nu_{\smf 2}$ term is the Laplacian viscosity and the others are
higher-order hyperdiffusivities. Specifically, $\bnabla^4 = \bnabla^2
\bnabla^2,$ $\bnabla^6 = \bnabla^2 \bnabla^2 \bnabla^2,$ and
$\bnabla^{[4]} = \partial_x^4 + \partial_y^4 + \partial_z^4.$
They differ in that
the higher the order, the more selective they are in diffusing high-k
structure while preserving low-k structure. Equivalently stated, high-order
hyperdiffusivities erase small-scale structure without affecting the larger
scales. For many physical applications, such as the
turbulent magnetic dynamo, the large scale structure is
unaffected by the nature of the small-scale diffusivity
(Haugen \& Brandenburg 2004), and so
in these instances, the use of hyperdiffusivity instead of Laplacian
diffusivity enhances the wavenumber resolution.

The $\bnabla^4$ operator is spherically symmetric in Fourier space
while the $\bnabla^{[4]}$ operator is not.  This affects the maximum-possible
timestep because in order to
be advectively stable, the Courant condition implies that the product $|k|
\Delta t$ must be less than a given value, and so the high-k corners of the 3D
Fourier cube are the most vulnerable to advective instability.
In these corners, the $\bnabla^4$ term delivers more diffusion
than the $\bnabla^{[4]}$ term, but with a
cost of twice as many finite difference operations.

Hyperdiffusivity acts together with high-order finite differences to
enhance the resolution of a simulation. High-order finite differences
allows structure to be finite differenced with less error,
and hyperdiffusivity allows this structure to evolve with less dissipation
than it would with Laplacian diffusivity.

Hyperdiffusivity can benefit the timestep as well as the resolution.  Suppose
one chooses the Laplacian term to provide the dominant fraction of the
diffusivity.  One typically then evaluates by trial and error the minimum value
of $\nu_{\smf 2}$ and the maximum timestep $\Delta t$ that one can get away
with to maintain stability. Once these values are chosen, the addition of a
small measure of hyperdiffusivity, small enough so as not to contribute
significantly to the total diffusivity, but large enough to affect the
highest-k structure, increases the maximum stable timestep by about 50\%.  This
technique works because the highest-wavenumber structure is the most vulnerable
to instability, and so the timestep depends most critically on the value of the
diffusivity for these wavenumbers.

One could additionally note that ``maximum stable timestep" is not sharply
defined. For instance, a timestep at the cusp of stability might be
unstable, but only after perhaps more than 10 timesteps. Also, the maximum
stable timestep depends on the maximum value of the velocity in the simulation,
which is changing in time.

\subsection{Magnetohydrodynamic equations}

The equations of incompressible MHD with diffusivity and hyperdiffusivity are
\begin{equation}
\partial_t \V = - \V \cdot \bnabla \V + \B \cdot \bnabla \B
+ \nu_{\smf 2} \bnabla^2 \V
- \nu_{\smf 4} \bnabla^4 \V \label{equtv}
- \nu_{\smf [4]} \bnabla^{[4]} \V \label{equtv}
+ \nu_{\smf 6} \bnabla^6 \V \label{equtv}
+ \nu_{\smf [6]} \bnabla^{[6]} \V \label{equtv}
+ \nu_{\smf D} \bnabla \bnabla \cdot  \V
\label{eqctv}
\end{equation}
\begin{equation}
\partial_t \B
=
\bnabla \times \left[
{\bf V} \times {\bf B}
- \eta_{\smf 2}     {\bf J}
+ \eta_{\smf 4}     \bnabla^2 {\bf J}
- \eta_{\smf 6}     \bnabla^4 {\bf J}
- \eta_{\smf 2,[4]} \bnabla^[4] {\bf J}
\bf \right]
+ \eta_{\smf D} \bnabla \bnabla \cdot \B
\label{eqctb}
\end{equation}
\begin{equation}
\J = \bnabla \times \B
\end{equation}

\begin{table}
\begin{center}
\begin{tabular}{
ll @{\hspace{25mm}}
ll}
\hline
\V         & Velocity   &  \B       & Magnetic field\\

              & &
\J & Current ($\bnabla \times \B$) \\

$\nu_{\smf 2}$    & Laplacian Viscosity   &
$\eta_{\smf 2}$ & Laplacian resistivity\\

$\nu_{\smf 4}$    & Hyperviscosity for $\bnabla^4$ &
$\eta_{\smf 4}$   & Hyperresistivity for $\bnabla^4$ \\

$\nu_{\smf [4]}$  & Hyperviscosity for $\bnabla^{[4]}$ &
$\eta_{\smf [4]}$ & Hyperresistivity for $\bnabla^{[4]}$ \\

$\nu_{\smf 6}$    & Hyperviscosity for $\bnabla^6$ &
$\eta_{\smf 6}$   & Hyperresistivity for $\bnabla^6 $ \\

$\nu_{\smf 2,[4]}$ & Hyperviscosity for $\bnabla^2 \bnabla^{[4]}$ &
$\eta_{\smf 2,[4]}$& Hyperresistivity for $\bnabla^2 \bnabla^{[4]}$ \\

$\nu_{\smf [6]}$  & Hyperviscosity for $\bnabla^{[6]}$ &
$\eta_{\smf [6]}$ & Hyperresistivity for $\bnabla^{[6]}$ \\

$\nu_{\smf D}$    & Divergence viscosity &
$\eta_{\smf D}$   & Divergence resistivity\\
\hline
\end{tabular}
\end{center}
\caption{ \label{tablevar}
Variables in the equations of MHD}.
\end{table}

The variables are defined in Table (\ref{tablevar}).  The magnetic
equation~(\ref{eqctb}) is written as a curl so that the contrained
transport scheme can preserve the magnetic divergence (Appendix).  The
remaining term proportional to $\eta_{\smf D}$ outside of the curl has
the effect of diffusing away any magnetic divergence present, and does
nothing otherwise (\S~\ref{divergence}).  Similarly, the term
proportional to $\nu_{\smf D}$ term can be used to help quell kinetic
divergence in incompressible simulations, although for the simulations
in this paper we use a spectral projection to accomplish this.

We clarify the meaning of equation~(\ref{eqctb}),
by assuming that $\bnabla \cdot \B = 0:$
\begin{displaymath}
\partial_t \B
= \B \cdot \bnabla \V - \V \cdot \bnabla \B - \B \, \bnabla \cdot \V
\end{displaymath}
\begin{equation}
\label{laplaceb}
+ \eta_{\smf 2}     \bnabla^2 \B
- \eta_{\smf 4}     \bnabla^4 \B
+ \eta_{\smf 6}     \bnabla^6 \B
+ \eta_{\smf 2,[4]} \bnabla^2 \bnabla^{[4]} \B
+ \eta_{\smf D}     \bnabla \bnabla \cdot \B
\end{equation}
The $\J$ terms in equation~(\ref{eqctb}) are seen to have the role of
diffusivities in equation~(\ref{laplaceb}). In the simulations in
\S~\ref{simulations}, we use equation~(\ref{eqctb}) for
constrained transport and equation~(\ref{laplaceb})
for unconstrained transport.

Time centering of the staggered-grid and centered-grid constrained
transport equations encounter a similar circumstance as with the
kinetic equation.  In the staggered-grid constrained transport
configuration employed by Hawley and Stone (1995), the electric fields
are spatially interpolated from the nearest 8 velocity and magnetic
field vectors, and they are also time-interpolated to the next half
step with an Alfv\'en wave Method of Characteristics scheme (Hawley
and Stone 1995).  The implicit diffusivity inherent in these
interpolations stabilizes the magnetic field evolution, even in the
absence of explicit diffusivity.  For Runge-Kutta timestepping
on a centered grid, there are no spatial interpolations in
constructing the electric field, and no accompanying intrinsic
diffusivity. Some form of diffusivity is then generally required to maintain
stability.

\subsection{Divergence} \label{divergence}

In a constrained transport simulation, the magnetic divergence remains
zero to machine precision. For unconstrained transport, we set the
magnetic divergence to zero with a Poisson projection in Fourier space
once every four timesteps. In tests, we found that the evolution is
virtually identical whether the divergence is removed once every
timestep or once every four timesteps (Maron 2004), and in both cases
the fractional magnetic divergence remains below one percent.  For
both constrained and unconstrained transport simulations of
incompressible flow that we describe below, the kinetic divergence is
removed once every timestep in Fourier space, although for
quasi-incompressible flow, it suffices to do this only once every few
timesteps (Maron 2004).

The divergence diffusivity term $\partial_t \B = \eta_{\smf D} \bnabla
\bnabla \cdot \B$ is helpful for reducing the effect of magnetic
divergence in the timesteps between Poisson projections.  It diffuses
any magnetic divergence present, but otherwise does nothing. If
each Fourier component has the form $\hat{\B}({\bf k}) =
\hat{\B}_\parallel {\bf k} / |{\bf k}| + \hat{\B}_\perp$, then the
$\bnabla \bnabla \cdot \B$ term evolves $\hat{\B}$ as $\partial_t
\hat{\B}_\parallel = - k^2 \hat{\B}_\parallel$ and $\partial_t
\hat{\B}_{\perp} = 0$.  The divergence diffusivity is also helpful for
a constrained transport simulation. Without it, CT preserves the
magnetic divergence but does nothing to remove it. With the divergence
diffusivity, any divergence present is diffused away.

\subsection{Initialization} \label{initial}

The constrained transport algorithm evolves the magnetic field in such
a way as to conserve magnetic divergence. If the initial conditions
have zero divergence, the divergence remains zero indefinitely.
Even if any monopoles do grow slowly, they can be removed at negligible
computational expense with a Fourier projection, say, once every thousand
timesteps.Care
must be taken with the initial conditions, however, because the
divergence depends on the method for approximately evaluating
derivatives. One may use, for example, a spectral or a finite
difference divergence operator. For constrained transport to work, the
derivatives used for making the initial conditions divergenceless must
be the same as those used in the simulation.  Even an analytic
function with vanishing divergence may not have vanishing numerical
divergence.

To initialize the magnetic field, we apply the following procedure.
In three dimensions, let the wavenumbers for the
magnetic field Fourier components $\hat{\B}$
be ${\bf k}$, the length of each side of the simulation volume be
${\bf L}$, and the number of grid zones on each side be ${\bf N}$. If
the magnetic divergence is defined spectrally, then the constraint on
the magnetic field is $\hat{\B}({\bf k}) \cdot {\bf k} = 0$.
For a finite difference derivative the constraint is slightly different.
If we use the high-order finite difference from \S~\ref{rez}
(eq.~\ref{stencil}), we can define an adjusted dimensionless wavenumber
by
\begin{equation}
k^*_i = \sum_{j=-S}^{S} m_j \sin(k_i L_i j / N_i),\end{equation} 
and then the finite difference condition for zero divergence
is
\begin{equation}\hat{\B}(\kk) \cdot \kk^* = 0
\label{divformula}.\end{equation} 
The initial conditions for a constrained
transport simulation should satisfy this condition, which is easily
implemented in Fourier space.

\subsection{High-wavenumber finite differences} \label{rez}

High-order spatial derivatives can enhance the wavenumber resolution of a
simulation. To quantify this, define a function $f_j(x_j)$ on a periodic grid
$x_j = j$, with $j$ an integer. Then construct a finite difference derivative
$f^\prime(0)$ at $x=0$ using a
radius-$S$ stencil. The familiar result for the gradient on
a radius-one stencil
is $f^\prime(0) \sim (f_{1} - f_{-1})/2$, which is obtained from fitting a
polynomial of order $2$ to $f_j$ at $j=0$. For an order 4 polynomial on a
radius-two stencil, \begin{equation}f^\prime(0) \sim \frac{1}{12} f_{-2}
  - \frac{2}{3} f_{-1} + \frac{2}{3} f_1 - \frac{1}{12} f_2.
\end{equation} 
For a stencil of order $S$,
\begin{equation} f^\prime(0) \sim \sum_{j=-S}^{S} m_j f_j \label{stencil}
\end{equation} where $m_{-j} = -m_j$. The coefficients for a radius-three stencil
are $\{m_1,m_2,m_3\}=\{3/4,3/20,1/60\}$.

\begin{figure}
\plotone{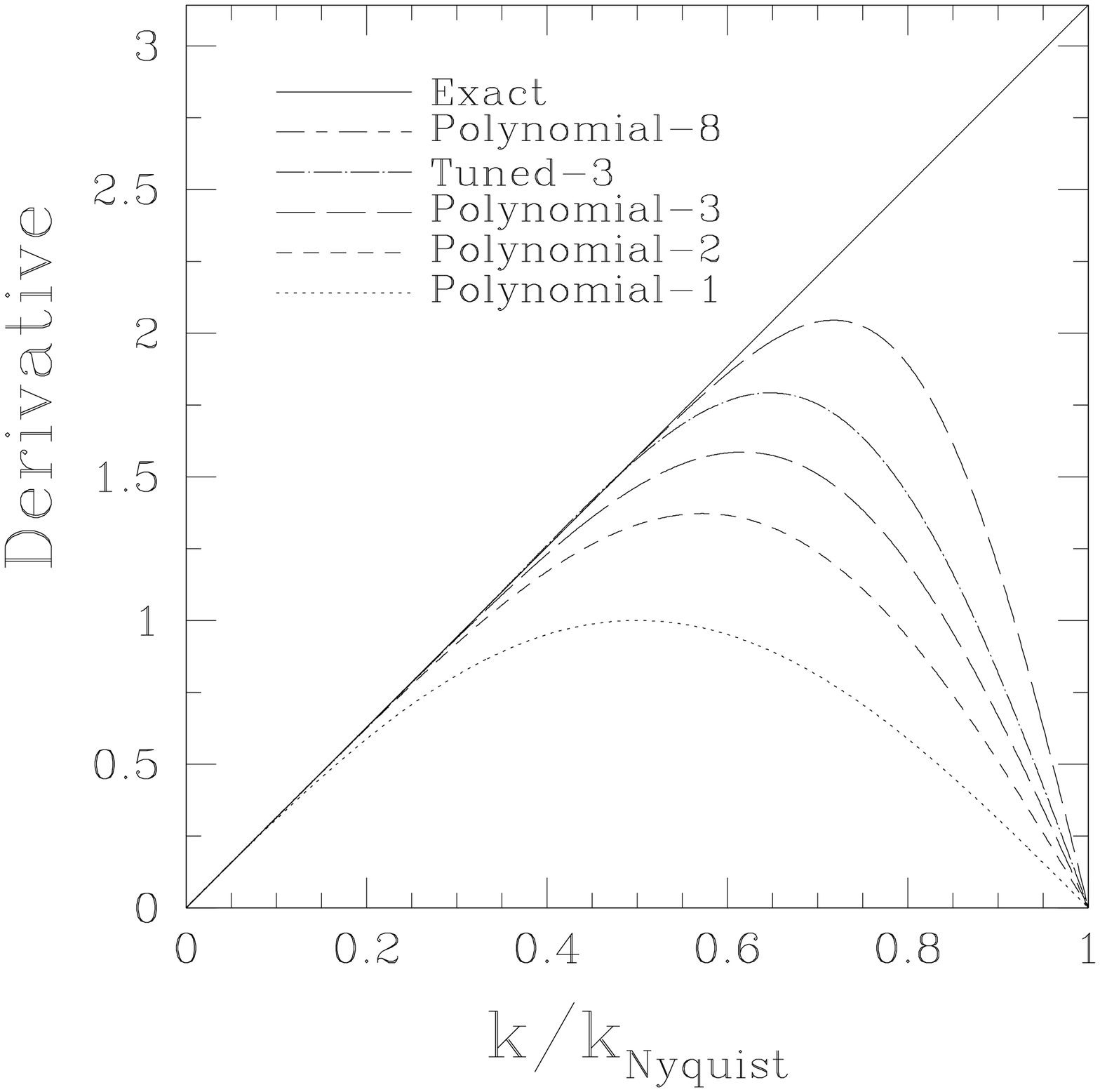}
\caption{ \label{figstencil}
The accuracy of finite-difference derivatives as a function of
the Nyquist-scaled wavenumber $k.$ ``Polynomial-N" denotes a
radius-N stencil with polynomial-based coefficients, and ``Tuned-N"
denotes a radius-N stencil with tuned coefficients (\S~\ref{rez} and
Table~\ref{tablek}). The ``Exact" curve is the derivative as
evaluated by a Fourier transform. A radius-N stencil with
polynomial-based coefficients is accurate to order $2N.$}
\end{figure}

Consider the finite-difference error at $x=0$
for a Fourier mode $\sin(\pi k x)$.
(Cosine modes can be ignored because they don't contribute to the derivative at
$x=0$.)  We scale the wavenumber $k$ to grid units so that $k=1$
corresponds to the maximum (Nyquist) wavenumber expressible on the grid. The
finite difference formula (eq.~\ref{stencil}) gives
\begin{equation}
f_k^\prime(0) \sim 2
\sum_{j=1}^S m_j \sin(\pi j k) \label{eqderiv}
\end{equation}
whereas the correct value is $k \pi$. The spectral procedure of taking
the derivative by transforming to Fourier space and back gives the
correct value up to $k=1$, but at a cost of $5 \log_2N$ floating point
operations per grid point per transform, where $N$ is the number of
grid points, whereas the finite difference derivative (equation
\ref{stencil}) on a radius $S$ stencil costs $3S-1$ floating point
operations.  Figure~(\ref{figstencil}) exhibits the accuracy of finite
difference derivatives of different orders as a function of
wavenumber.  The wavenumber resolution increases with polynomial order.

The polynomial fit can be tuned by adjusting the coefficients to
enhance high-wavenumber accuracy substantially at the expense of a
negligible loss of accuracy at low wavenumbers (Maron
2004). Table~\ref{tablek} gives the coefficients for finite difference
operators at various orders, along with the maximum wavenumber for
which they are 0.5\% accurate.  In Figure~\ref{figstencil} we also
show the wavenumber accuracy for a
radius-three stencil using tuned coefficients,
designed to have a relative precision of $< 0.5$\% for $k=0$ to $0.50$. We
will see in \S~\ref{overlap} that a simulation based on this tuned radius-three
scheme performs better than a radius-four simulation with conventional
coefficients.
We note that power spectra do not reveal a difference
between turbulence simulations with tuned and untuned coefficients,
because errors in the derivative manifest as an advective dispersion
rather than as a diffusivity.  One has instead to examine the fields
in real space rather than in Fourier space.

\begin{table}
\begin{center}
\begin{tabular}{lrrrrrrrrrl}
\hline
Operation &\bs $m_0$&\bs $m_1$&\bs $m_2$&\bs $m_3$&\bs $m_4$&\bs $m_5$&
\bs $m_6$&$m_7$&$m_8$ & $k_{max}$\\
\hline

$\partial/\partial x$\,(P1)&\bs 0&\bs 0.50000&\bs &\bs &
\bs &\bs &\bs &\bs &\bs & 0.12\\

$\partial/\partial x$\,(P2)&\bs 0&\bs 0.66667&\bs -0.08333 & \bs &
\bs &\bs &\bs &\bs &\bs & 0.24\\

$\partial/\partial x$\,(P3)&\bs 0&\bs 0.75000&\bs -0.15000 &\bs 0.01667&
\bs &\bs &\bs &\bs &\bs & 0.34\\

$\partial/\partial x$\,(P4)&\bs 0&\bs 0.80000&\bs -0.20000 &\bs 0.03810&
\bs-0.00357 &\bs &\bs &\bs &\bs & 0.40\\

$\partial/\partial x$\,(P5)&\bs 0&\bs 0.83333&\bs -0.23810 &\bs 0.05952&
\bs-0.00992 &\bs 0.00079&\bs &\bs &\bs & 0.44 \\

$\partial/\partial x$\,(P8)&\bs 0&\bs 0.88889&\bs -0.31111&\bs 0.11313&
\bs -0.03535&\bs 0.00870&\bs -0.00155&\bs 0.00018&\bs -0.00001 & 0.56 \\

$\partial/\partial x$ (T3)& \bs 0 &\bs 0.81796&\bs -0.21324&\bs 0.03683&
\bs        &\bs &\bs &\bs &\bs & 0.50 \\

$\partial/\partial x$ (T8)& \bs 0 &\bs  0.95951 &\bs -0.42312 &\bs 0.22746 &
\bs -0.12450 &\bs 0.06461 &\bs -0.03004 &\bs 0.01156 &\bs -0.00273 & 0.76\\


$\partial/\partial x$ (T8)& \bs 0 &\bs  0.96685    &\bs -0.43635  &\bs  0.24410  &
\bs -0.14168       &\bs  0.07976 &\bs -0.04148 &\bs  0.01884 &\bs -0.00536 & 0.80 \\


$\partial^2/\partial x^2$ \bss (P3)&\bs -2.72222&\bs 1.5&\bs -0.15&
\bs 0.01111&\bs &\bs &\bs &\bs &\bs & \\

$\partial^4/\partial x^4$ \bss (P3)&\bs 9.33333 &\bs -6.5 &\bs 2.&
\bs -0.16667&\bs &\bs &\bs &\bs &\bs & \\

$\partial^6/\partial x^6$ \bss (P3)&\bs -20.&\bs 15. &\bs -6. &
\bs 1. &\bs &\bs &\bs &\bs &\bs & \\

Interpolation \bss (P3) & \bs 0 &\bs  0.58594 &\bs -0.09766 &\bs  0.01172 &
\bs        &\bs &\bs &\bs &\bs & 0.35 \\

Interpolation \bss (T3) & \bs 0 &\bs  0.60103 &\bs -0.12312 &\bs  0.02214 &
\bs        &\bs &\bs &\bs &\bs & 0.50 \\

Interpolation \bss (T8) & \bs 0 &\bs   0.63099 &\bs -0.19580 &\bs  0.10149&
\bs  -0.05770 &\bs  0.03248&\bs -0.01709&\bs  0.00793&\bs -0.00233& 0.80  \\


\hline
\end{tabular}
\end{center}
\caption{ \label{tablek}
Coefficients for finite difference
operations. The column $``k_{max}"$ indicates the maximum wavenumber
for which the operator is no worse than $0.5 \%$ accurate.  For the odd-order
derivatives, $m_{-j} = -m_j$, and for the even-order derivatives as
well as the interpolations, $m_{-j} = m_j$. The interpolation
coefficients are for interpolation to a point halfway between two grid
zones using the nearest $S$ grid points on each side (\S~\ref{interpolate})
}.  
\end{table}

\subsection{Dealiasing} \label{dealiasing}

Nonlinear terms in the MHD equations such as $\V \cdot \bnabla \V$
suffer an aliasing error for high-wavenumber structure.  We first
illustrate this with a 1D example before treating the 3D case (Canuto
et al.\ 1987).  Let $A_j$ and $B_j$ be discrete functions on a
one-dimensional grid with $N=16$ grid points: $1 \leq j \leq N$, and
with accompanying Fourier modes ranging from $-N/2 \leq s \leq
(N/2)-1.$ The Fourier expansion of $A$ is \begin{equation} A_j =
\sum_{s=-N/2}^{N/2-1} \hat{A}_s \, \exp(2 \pi i j s / N)
\end{equation} and similarly for $B_j$. Let $A$ and $B$ each be
composed of a single Fourier mode with $s=6$. The product $C = A B$ is
then composed of a single Fourier mode with $s=12$, but in this
discrete representation, this is equivalent to $s=-4$. This remapping
of high wavenumber modes to low wavenumber modes is known as aliasing
error.  Note that $s=6$ is within the Nyquist limit of $|s| \leq N/2$
while the product mode $s=12$ is not. While $A_j$ and $B_j$ might be
resolvable on the grid, their product need not be.  This can be fixed
by truncating before and after the product all modes outside $|s| \leq
N/3$ to ensure that no modes in the product will exceed the Nyquist
limit.  A spectral code can make this truncation because the fields
are in Fourier space, while a finite difference code cannot. As a
practical matter it suffices to set the diffusivities high enough so
that negligible structure exists outside the $N/3$ aliasing limit.

Alternatively, 
the aliasing problem can be remedied with a
staggered-grid correction (Canuto et al.\ 1987), and then
the truncation isn't necessary.  Additionally, this allows
us to simulate structure beyond $k=2/3.$ We first exhibit this
procedure in 1D and then extend it to 3D below. To implement
the staggered-grid correction, first construct the
usual product $C_j = A_j B_j$. Then interpolate the centered grids
$A_j$ and $B_j$ to the staggered grids $A_{j+1/2}$ and $B_{j+1/2}$,
multiply to produce $C_{j+1/2}$ and transform back to the centered
grid $C_j^\prime$. Finally, combine the centered and staggered results
as $C_j \leftarrow (C_j + C_j^\prime) / 2$. This result is free from
aliasing error for all Fourier modes, and so no Fourier-space
truncation is necessary.  In Fourier space, the fields on the
staggered grid can be computed exactly by applying a phase shift to
each Fourier mode. A finite difference scheme can accomplish this with
the high-order interpolation discussed in \S~\ref{stagger}.

In three dimensions the aliasing error can be eliminated by truncating
the Fourier modes outside $|{\bf \vec{s}}| < {\bf \vec{N}}/3$, where
${\bf \vec{s}} = \{s_x,s_y,s_z\}$ and ${\bf \vec{N}} = \{N_x, N_y,
N_z\}$. The grid-shift correction is more complicated. Seven grid
shifts are needed to completely correct the error, but it is more
efficient to settle for a more limited correction involving just one
shift (Canuto et al.\ 1987).  Interpolate $A$ and $B$ to the staggered
grid $\{j_x+1/2,j_y+1/2,j_z+1/2\}$, multiply, return to the centered
grid $\{j_x,j_y,j_z\}$, and as in the one-dimensional case, average
the result with the product carried out on the original centered grid.
This yields an alias-free result if accompanied by a Fourier
truncation of all modes outside $|{\bf \vec{s}}| < \sqrt{3} {\bf
\vec{N}}/2$, which is 94 \% of the Nyquist limit. For a finite
difference code, the diffusivity can be set to minimize the energy
outside the truncation zone.
When the three-dimensional, staggered-grid, aliasing correction is
combined with high-order, volume-centered, constrained transport,
magnetic divergence is still preserved to machine precision. This
scheme uses the high-order finite differences and interpolations
discussed in \S~\ref{rez} and \S~\ref{interpolate}.

\subsection{Interpolation} \label{interpolate} \label{stagger}

High-order interpolation allows implementation of the staggered
aliasing correction discussed in \S~\ref{dealiasing}.  It can also be
used for doubling the grid size. In this case, the new points in the
doubled three-dimensional grid can be generated with a set of
interpolations along and diagonal to the three grid axes. Doubling the
grid is also useful for interpolating to points that are not exactly
halfway between grid points. For example, the grid can be doubled with
the high-order interpolation discussed above, and then a simpler
algorithm can be used to further interpolate to a point anywhere
within the refined grid.  We used this technique to trace the path of
magnetic fieldlines in our studies of cosmic ray diffusion (Maron
2003), and also for doubling the resolution of turbulent dynamo
simulations in Maron et al.\ (2004).  Lastly, we have found that
high-order tuned derivatives and interpolations are useful for
post-simulation analysis of timeslice data without the need for
computationally expensive fast Fourier transforms.

A high-order interpolation to a point halfway between two grid points
can be accomplished with
\begin{equation}
f_{j+1/2} = \sum_{i=1}^S m_i (f_{j+i} + f_{j+1-i}).
\label{eqinterp}
\end{equation}
The error in the interpolation as a function of wavenumber can be calculated by
applying
equation~(\ref{eqinterp}) to a cosine mode centered on $f_{j+1/2}$:
$f_i = \cos(\pi k (i - (j + 1/2)))$.
The interpolated value is
\begin{equation}
f_{j+1/2}^* = \sum_{i=1}^S m_j \cos(\pi i k)
\end{equation}
whereas the correct value is $f_{j+1/2} = 1$.
Table~\ref{tablek} gives coefficients for
an interpolation based on a radius-three
stencil from a polynomial fit to $f_j$, in the row labeled
``Interpolation (P3)." Tuning the coefficients can improve the
wavenumber resolution of the interpolation in a similar manner as was done
for the derivative.
Tuned
coefficients for radius 3 and 8 stencils are given in Table~(\ref{tablek})\,
in the rows labeled T3 and T8, with the column labeled $k_{max}$
giving the maximum wavenumber for which the interpolation is $0.5 \%$ accurate.

\subsection{Operation count} \label{sectionop}

A finite difference derivative on a radius $S$ stencil costs $S$ multiplies and
$2S - 1$ adds. Since add and multiply units tend to come in pairs on most
modern machines, this is effectively $4S - 2$ floating point operations per
grid cell. A code with $F$ finite difference convolutions per timestep involves
$F(4S-2)$ floating point operations.

The computation cost per timestep scales as the number of derivatives
computed per grid element per time update. For constrained transport,
one needs to calculate three derivatives for $\partial_i \rho$, three
for $\partial_i e$, nine for $\partial_i \V_j$, six for $\bnabla
\times \B$ and six for $\bnabla \times \E$. Diffusion terms such as
$\bnabla^2 \V$ and $\bnabla^2 \J$ aren't included in the tally because
they need only be applied every few timesteps (Maron 2004). The same
is true for the Fourier-space Poisson projection to make the fields
divergenceless. Maron (2004) found that the results were
effectively identical whether these terms were applied once every
timestep or once every four timesteps.

In Table~\ref{tableop} we list the number of derivatives computed per timestep
for various classes of algorithms.  Pencil is a vector potential code
(Brandenburg \& Dobler 2002) that calculates $\J$ directly from $\A$ to take
advantage of the cache memory.  This involves 15 distinct 2-level derivatives
computed from a 2D stencil.
For a radius-S finite difference, a 2D derivative
involves $4S^2 - 1$ adds,
compared to $2S - 1$ adds for a conventional 1D derivative.
We don't consider multiples because they are always less numerous than
adds, and because CPUs tend to feature
add and multiply units in pairs.
As a result, a 2D derivative corresponds to $2S+1$ times the computational
effort of a conventional 1D derivative.
Thus, the calculation of $\J$ with $S=3$ corresponds effectively to about $105$
conventional derivatives.
Alternatively, a vector potential algorithm can instead calculates
$\B$ from $\A$ and then $\J$ from $\B$ in separate stages, at a possible
cost of extra memory latency. In table~\ref{tableop}, we denote a vector
potential algorithm that calculates $\J$ directly from $\A$ as ``1-stage,"
and a vector potential algorithm that calculates $\B$ from $\A$ and then
$\J$ from $\B$ as ``2-stage."
The constrained transport
algorithm evolves $\V$ and $\B$ according to equations~(\ref{eqctv})
and~(\ref{eqctb}), and the unconstrained transport algorithm evolves $\V$ and
$\B$ according to equations~(\ref{eqctv}) and~(\ref{laplaceb}). The spectral
algorithm computes 15 Fourier transforms instead of derivatives and thus
doesn't directly compare with the finite difference codes.

A vector potential code such as Pencil can be adapted to the algorithm
described here by substituting equation~(\ref{eqctb}) for the vector
potential equation, and by changing the finite difference coefficients
from the polynomial-based values to the tuned values. This allows one to
compute boundary conditions using the magnetic field instead of the
vector potential.  However, there is an extra round of inter-processor
communication associated with calculating ${\bf J}$ from $\B$ and then
applying the curl.

\begin{table}
\begin{center}
\begin{tabular}{lll}
\hline
Algorithm & Terms & Derivatives \\
\hline

Constrained transport &
$\partial_i \rho$, $\partial_i e$, $\partial_i \V_j$,
$\bnabla \times \B$, $\bnabla \times \E$&
27\\

Unconstrained transport &
$\partial_i \rho$, $\partial_i e$, $\partial_i \V_j$, $\partial_i \B_j$,
$\partial_i (\bnabla \cdot \B)$ &
30 \\

2-stage vector potential &
$\partial_i \rho$, $\partial_i e$, $\partial_i \V_j$,
$\bnabla \times \A$, $\bnabla \times \B$ &
27 \\

1-stage vector potential (Pencil) &
$\partial_i \rho$, $\partial_i e$, $\partial_i \V_j$,
$\bnabla \times \A$, $\bnabla \times \bnabla \times \A$ &
120 \\


Spectral &
15 Fourier transforms &
N/A \\


\hline
\end{tabular}
\end{center}
\caption{
\label{tableop} The number of derivatives computed per time update
(see \S~\ref{sectionop}).}
\end{table}

A convenient unit for execution speed is
Kilo-grid Elements per CPU GigaFLOPS per Second (KEGS).  A constrained
transport code with a radius-three stencil and two Runge-Kutta steps per
timestep
would run ideally at a speed of 1850 KEGS. The actual speed is slower of course
because of overhead and because the finite difference convolutions don't run at
the peak floating point speed.  In benchmarks, we have observed that
the Pencil code, which has 3 Runge-Kutta
stages, runs at a speed of around 100 KEGS for serial operation and

50 KEGS for massively parallel operation.  For comparison, the highly-optimized
spectral-MHD code Tulku (Maron \& Goldreich 2001), has a speed of around
80 KEGS in serial and 40 KEGS in parallel (Maron 2004).

\section{Simulations} \label{simulations}

\subsection{The turbulent nonhelical dynamo} \label{dynamo}

\begin{figure}
{\plottwo{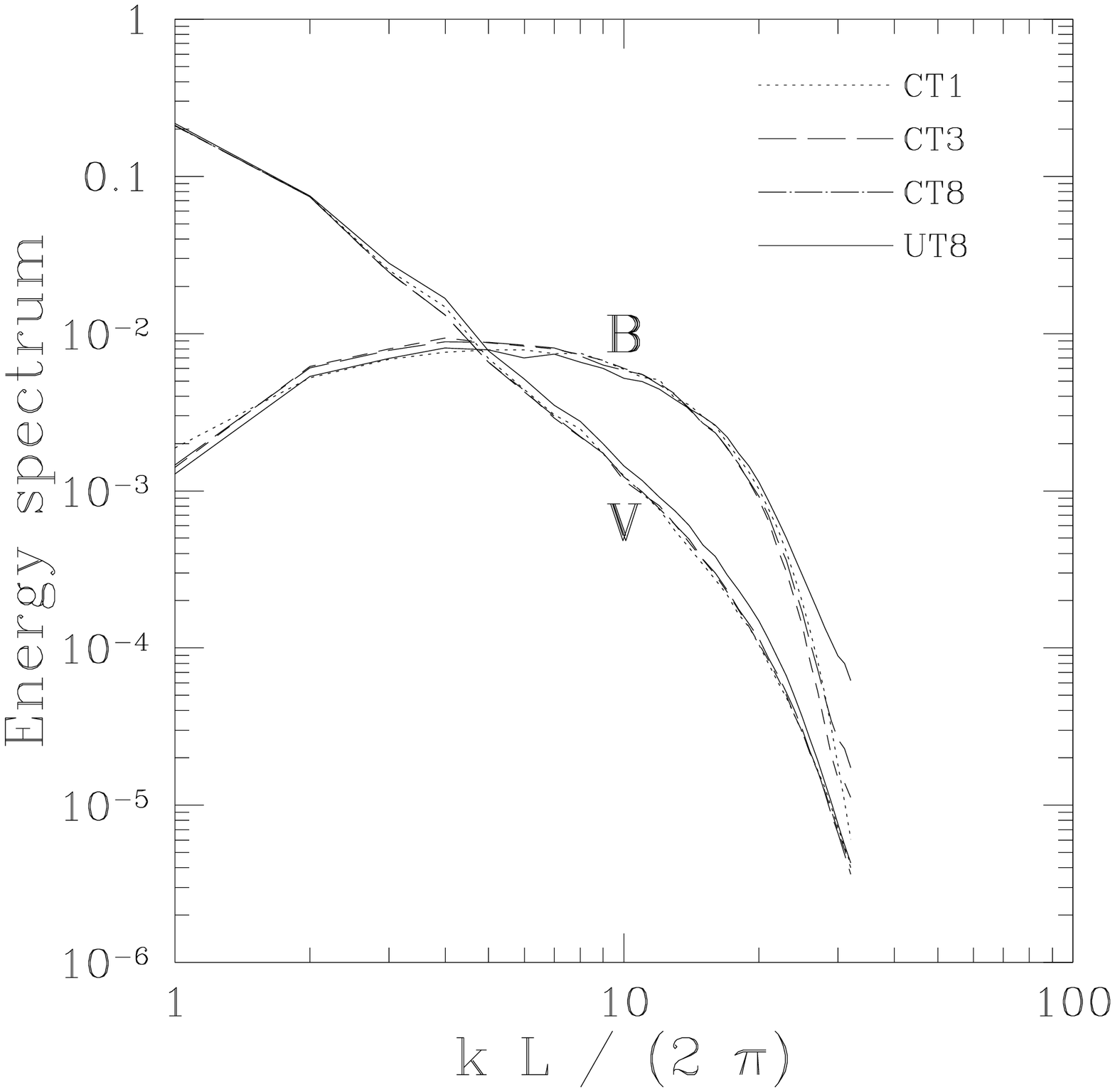}{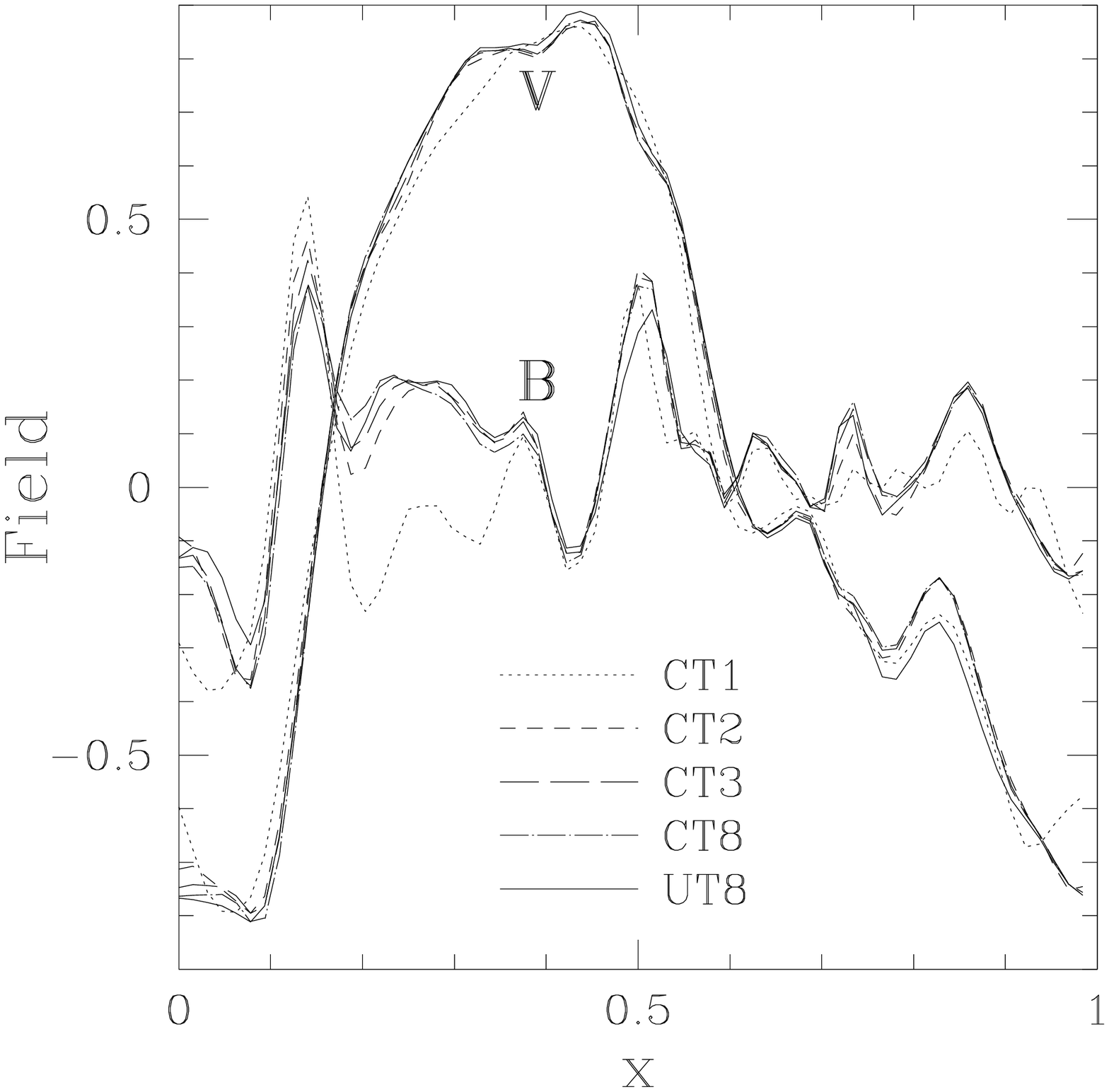}}
\caption
{ \label{figct} {\em (a)} The kinetic ($\V$) and magnetic ($\B$)
spectra for a set of simulations with identical initial conditions,
after two crossing times.  {\em (b)} The value of $B_y$ along a grid
line parallel to the $x$-axis after 0.4 crossing times, for the same
set of simulations.  The high-order constrained and unconstrained
transport simulations (CT3, CT8 \& UT8) more closely resemble each
other than they do the low-order constrained transport simulations
(CT1 \& CT2).}
\end{figure}

Our first test model is the turbulent, nonhelical, MHD dynamo
(Batchelor 1950), which is the magnetic analogue of the Kolmogorov
cascade for hydrodynamic turbulence.  The Kolmogorov cascade is the
long-term, steady state of hydrodynamic turbulence that is forced at
the large scale and dissipated by viscosity at the small scale, and it
has an energy spectrum of $E(k) \sim k^{-5/3}$. The nonhelical MHD
dynamo has the same setup but includes a spatially homogeneous
magnetic field with unit magnitude and zero mean. Maron et al.\ (2004)
found that the steady state of this system has a kinetic
energy spectrum of $E(k) \propto k^{-2}$, where the energy is
dominantly in large-scale eddys, and a magnetic spectrum with the
energy predominantly in the smallest-scale (resistive-scale) magnetic
structures. The kinetic and magnetic spectra are shown in figure
\ref{figct}.

\subsubsection{Models}

We ran a set of simulations of forced homogeneous MHD turbulence (Table
\ref{tablesim}) with and without constrained transport and at various spatial
orders to test the effectiveness of high-order constrained
transport. Each simulation has the same timestep,
viscosity and resistivity. The grid used in all cases
has $64^3$ zones covering a periodic unit cube.  The forcing power, density,
initial RMS velocity and magnetic field are all unity, and they remain so in
the long-term steady state.  

The initial state is the same for all
simulations. It was taken from a simulation of forced magnetized turbulence
(\S~\ref{dynamo}) in the long-term steady state. The magnetic divergence is
zero in Fourier space: $\hat{\B}(\kk) \cdot \kk = 0$. The initial state was
modified slightly for the constrained transport simulations to make it
magnetically divergenceless according to the finite-difference derivative
(\S~\ref{initial}).  This amounts to a very small adjustment in the fields,
much less than the difference between the fields after a crossing time of
evolution.

Constrained transport needs some form of diffusivity, either Laplacian
or hyperdiffusivity, to maintain stability. With diffusivity, it is
stable indefinitely, whereas it is unstable without it. To establish
this, we evolved the MHD equations with constrained transport for ten
crossing times stably.  The diffusivities used in these models are
Laplacian viscosity $\nu_2 = 10^{-3}$, hyperviscosity $\nu_{\smf [4]} = 2.5
\times 10^{-8}$ and hyperresistivity $\eta_{\smf 2,[4]} = 7 \times 10^{-12}$,
which we empirically find are the minimum values that maintain
stability for $V_{\smf RMS} = B_{\smf RMS} = 1$ on a $64^3$ grid.
Expressed in dimensionless form:
$\overline{\nu}_n = \nu_n / (V_{\smf RMS} dx^{n-1})$
and $\overline{\eta}_n = \eta_n / (V_{\smf RMS} dx^{n-1}),$ these are
$\overline{\nu}_2 = .064,$ $\overline{\eta}_{\smf [4]} = 6.6 \cdot 10^{-3}$
and $\overline{\eta}_{\smf 2,[4]} = 7.5 \cdot 10^{-3}.$

\begin{table}
\begin{center}
\begin{tabular}{|llrlll|}
\hline
Tag   & Method  & Stencil & Finite difference & $k_{max}$ & $L_2$ norm\\
      &         & radius  & technique         &           & \\
\hline
CT1   & Constrained transport   & 1 & Polynomial           & 0.12 & 0.561 \\
CT2   & Constrained transport   & 2 & Polynomial           & 0.24 & 0.193 \\
CT3   & Constrained transport   & 3 & Polynomial           & 0.34 & 0.083 \\
CT4   & Constrained transport   & 4 & Polynomial           & 0.40 & 0.042 \\
CT5   & Constrained transport   & 5 & Polynomial           & 0.44 & 0.023 \\
CT8   & Constrained transport   & 8 & Polynomial           & 0.56 &  0   \\
CT3t1 & Constrained transport   & 3 & Tuned                & 0.50 & 0.034 \\
CT3t2 & Constrained transport   & 3 & Tuned                & 0.52 & 0.037 \\
UT8   & Unconstrained transport & 8 & Polynomial           & 0.56 & 0.267 \\
\hline

\end{tabular}
\end{center}
\caption{
\label{tablesim}
Index of simulations. CT and UT denote constrained
and unconstrained transport. The column labeled ``finite difference technique"
specifies whether
the finite difference coefficients are from a polynomial fit or if they
have been tuned to enhance high-wavenumber accuracy
(\S~\ref{rez}).
$k_{max}$ denotes the largest wavenumber for which derivatives are $0.5 \%$
accurate.
The $L_2$ norms are taken with respect to UT8, as discussed in \S~\ref{overlap}
The simulations with tuned coefficients are
CT3t1, which has $k_{max}$
set to $0.50$, and CT3t2, which has $k_{max}$ set
to $0.52$.
}
\end{table}

The forcing is the same as used by Maron et al.\ (2004).  A random forcing
field is added to the velocity every timestep.  The spectrum of the forcing
field is $k^{-5/3}$, truncated 2.5 lattice units from the origin in Fourier
space, and the Fourier components have random phases. The forcing power,
simulation volume and density are unity, which yields RMS velocity and
magnetic fields of order unity (Maron 2004).

\subsubsection{Results} \label{overlap}

The diffusivities are
given in \S~\ref{simulations}.
We plot the kinetic and magnetic spectra in figure \ref{figct}{\em (a)}.  The
spectra are very similar for constrained and unconstrained transport
simulations, and also for different orders.  However the spectra alone
do not distinguish between simulations of different orders because an
error in the derivative manifests itself as an advective dispersion
rather than as a diffusivity (\S~\ref{rez}). One instead has to
examine the fields in real space.

In Figure~\ref{figct}{\em (b)}, we compare the magnetic fields at
t=0.4 crossing times. For the comparison, we examine the difference
between the fields integrated over space by computing the $L_2$ norm
between simulations $i$ and $j:$ \begin{equation} A_{ij}^2 = \frac{
\int [B_y(i) - B_y(j)]^2 d(\mbox{Vol})} {\int B_y(i)^2 d\mbox{Vol} +
\int B_y(j)^2 d(\mbox{Vol})}.
\end{equation} 
Stone et. al. (1992b) argue that this kind of comparison is more
meaningful than merely plotting the overlay of both fields.  The
constrained transport simulation with polynomial-based finite
differences on a radius-eight stencil (CT8 in Table \ref{tablesim}) serves
as the basis of comparison.
We compare the constrained transport simulations to an unconstrained
transport simulation on a radius-eight tuned finite-difference stencil (UT8).
We use UT8 as a stand-in for the spectral algorithm because of its high
wavenumber resolution.

The spectral algorithm delivers the highest-attainable resolution because
spectral derivatives are exact for all wavenumbers.  With this, a 3D
spectral simulation without an aliasing grid-shift correction can resolve
structure up to k=2/3, and with a grid-shift correction it can resolve up to
k=.94 (Canuto 1987).  The spectral algorithm can also set the magnetic
divergence to zero in Fourier space at negligible cost.
Unconstrained transport does not explicitly conserve
magnetic divergence and so in model UT8 the divergence is cleaned with a
Fourier projection every timestep. We also tried applying the correction
every fourth timestep and with virtually identical results.
The radius-eight stencil of UT8 yields derivatives that are accurate up to
k=.56.

The $L_2$ norms given in
Table~(\ref{tablek}) show how the simulations progressively approach
the CT8 result as the stencil size increases.  
The match is poor for CT1 and better for CT3. 
We also note that the radius-three simulation with
tuned derivatives (CT3t) performs better than the radius-four simulation
with polynomial-based derivatives, establishing the effectiveness of
tuned derivatives.
This can also be qualitatively seen in Figure (\ref{figct}), where
we see that the fields for CT8 and UT8 are closely aligned
(Fig.~\ref{figct}), and that they also
closely resemble those for CT3.

We attribute the remaining differences between CT8 and UT8 to the fact that
the magnetic divergence is removed spectrally in UT8, while it is handled
by constrained transport in CT8.
Collectively, the high-order constrained and unconstrained transport
simulations (CT3, CT8 \& UT8) more closely resemble each other than they do the
low-order constrained transport simulations (CT1 \& CT2).  We conclude that CT3
is already a good approximation to the spectral algorithm.

\subsection{Comparison of the constrained transport and vector potential
techniques} \label{pencil}

\begin{figure}
\plottwo{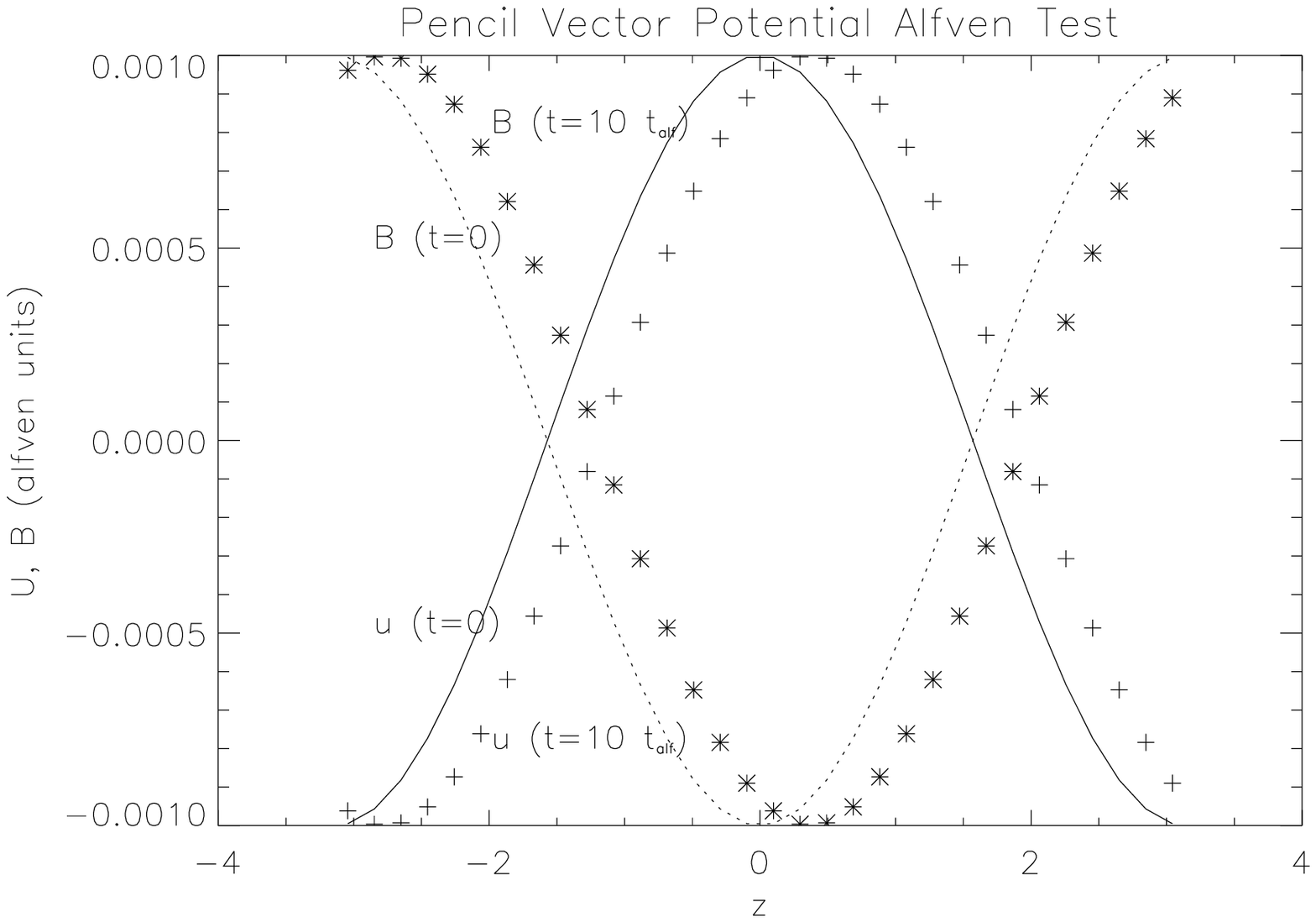}{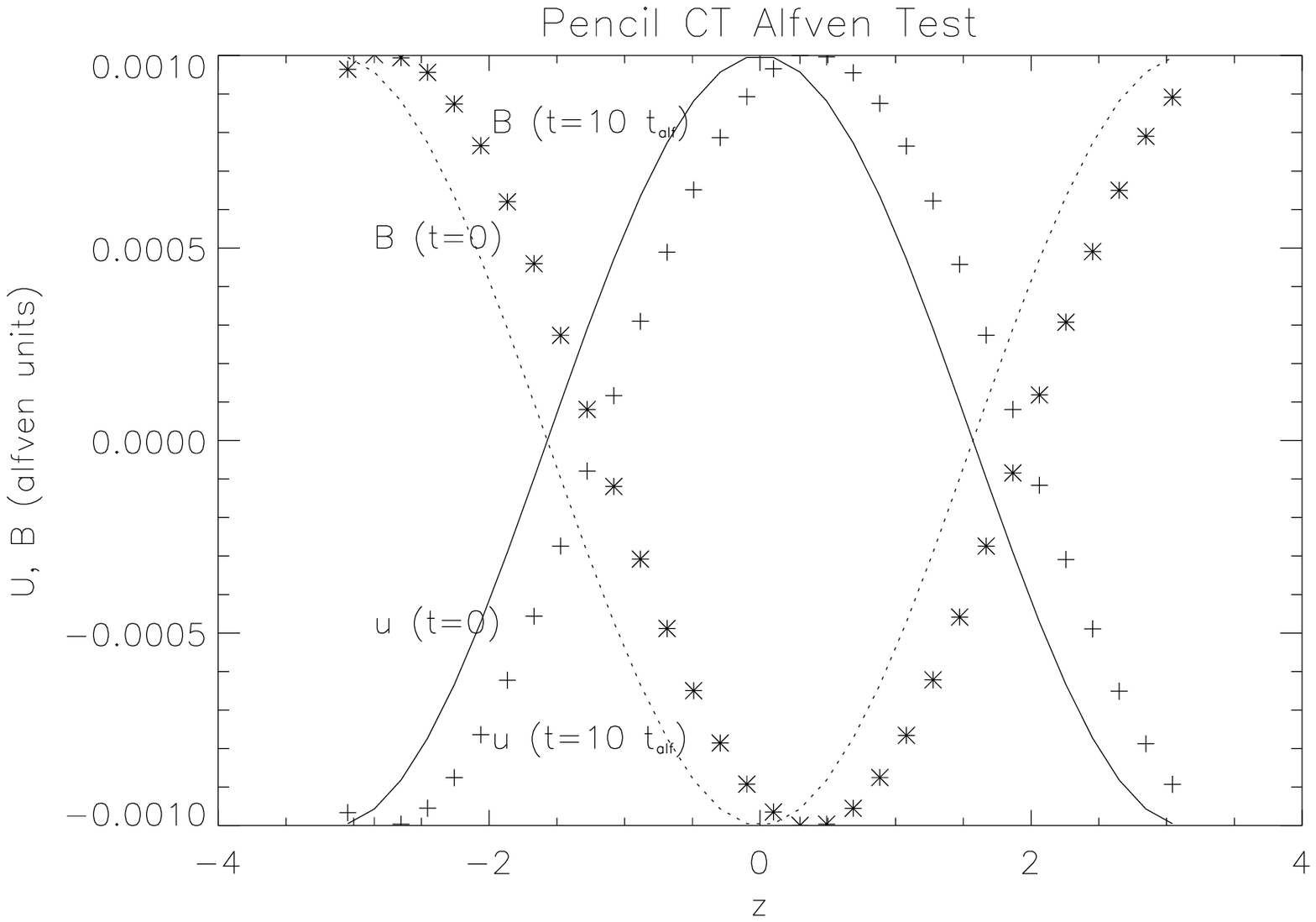}
\caption{
\label{figalfven} The Alfv\'en wave profile after ten crossing
times, with {\em (a)} the Pencil code using the vector potential form
of the induction equation and {\em (b)} using CT (\S~\ref{pencil}).
Alfven units denote the fluctuating field strengths compared to the
uniform component of the magnetic field.
} \end{figure}

We adapted the vector potential code Pencil to run in
CT mode and used it to compare the vector potential and CT techniques.
We ran an Alfv\'en wave
on a $32^3$ grid with zero viscosity and resistivity.
(Figure~\ref{figalfven}).
After ten crossing times, both the vector potential and CT techniques yield
wave profiles that agree with each other to within 1 percent.
The shape of the profiles are also well-matched with the initial conditions,
with a phase error of 10 percent.

\begin{figure}
\plotone{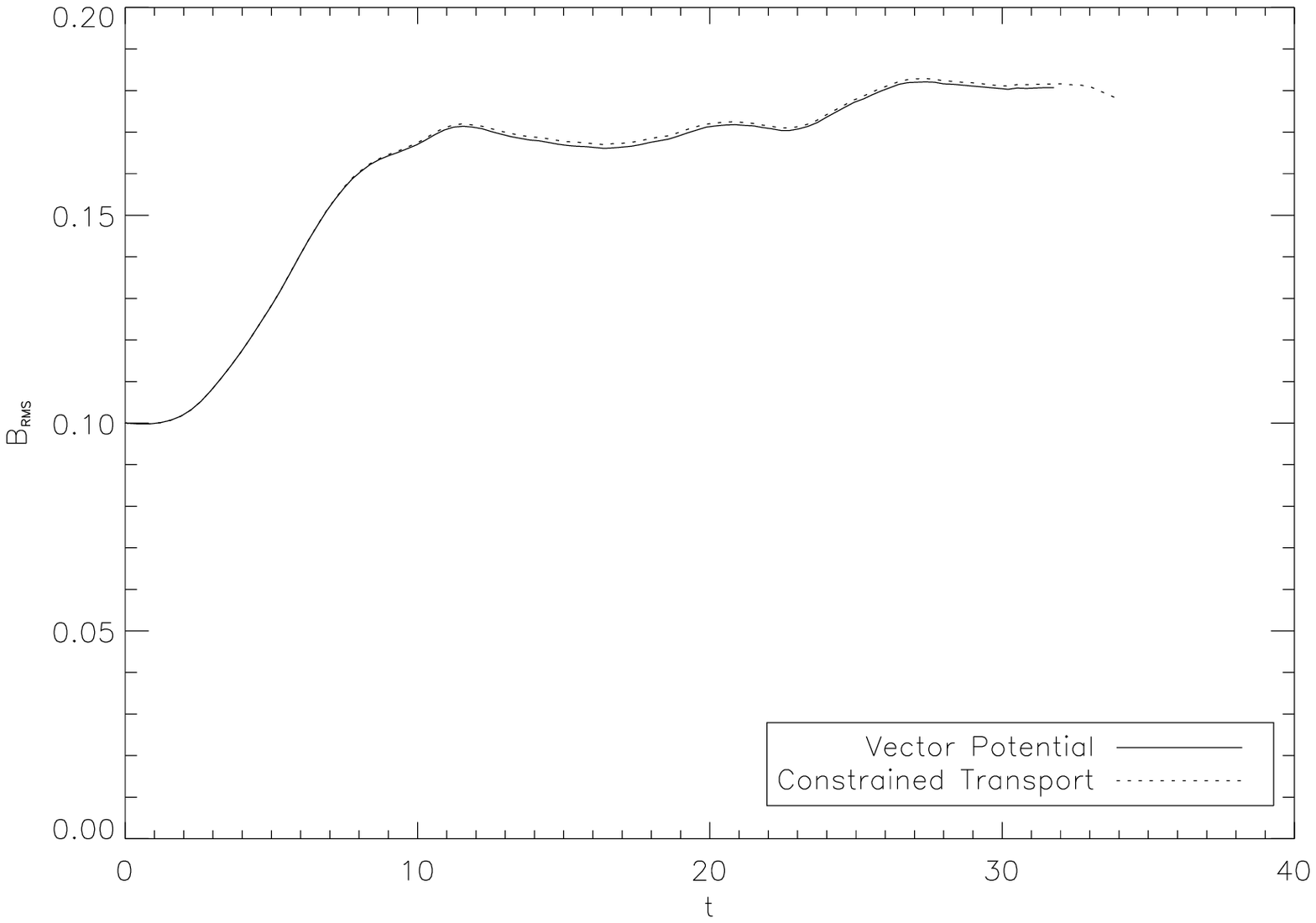}
\caption{
\label{figdynamo} We plot the RMS magnetic field strength for a
turbulent dynamo simulation using the Pencil code
in vector potential and CT mode (\S~\ref{pencil}).
Both techniques agree to 1 percent after 30 crossing times.
}
\end{figure}

We also used both techniques to run a turbulent dynamo simulation 
We started with an initially
weak magnetic field in the form of a Beltrami wave and applied helical forcing
until it grew to a steady state. The box size is $(2\pi)^3$, the density
is unity, the forcing power is equal to $0.07,$
the viscosity is equal to $5 \cdot 10^{-3},$ and the resistivity is equal
to $5 \cdot 10^{-3}.$
The RMS magnetic field
strength is plotted in Figure~(\ref{figdynamo}). 
After 30 crossing times, the values for $B_{RMS}$ for the CT and vector
potential techniques agree to 1 percent (Figure~\ref{figdynamo}).

\subsection{Oblique Alfv\'en wave test} \label{sectionalfven}

\begin{figure}
\plotone{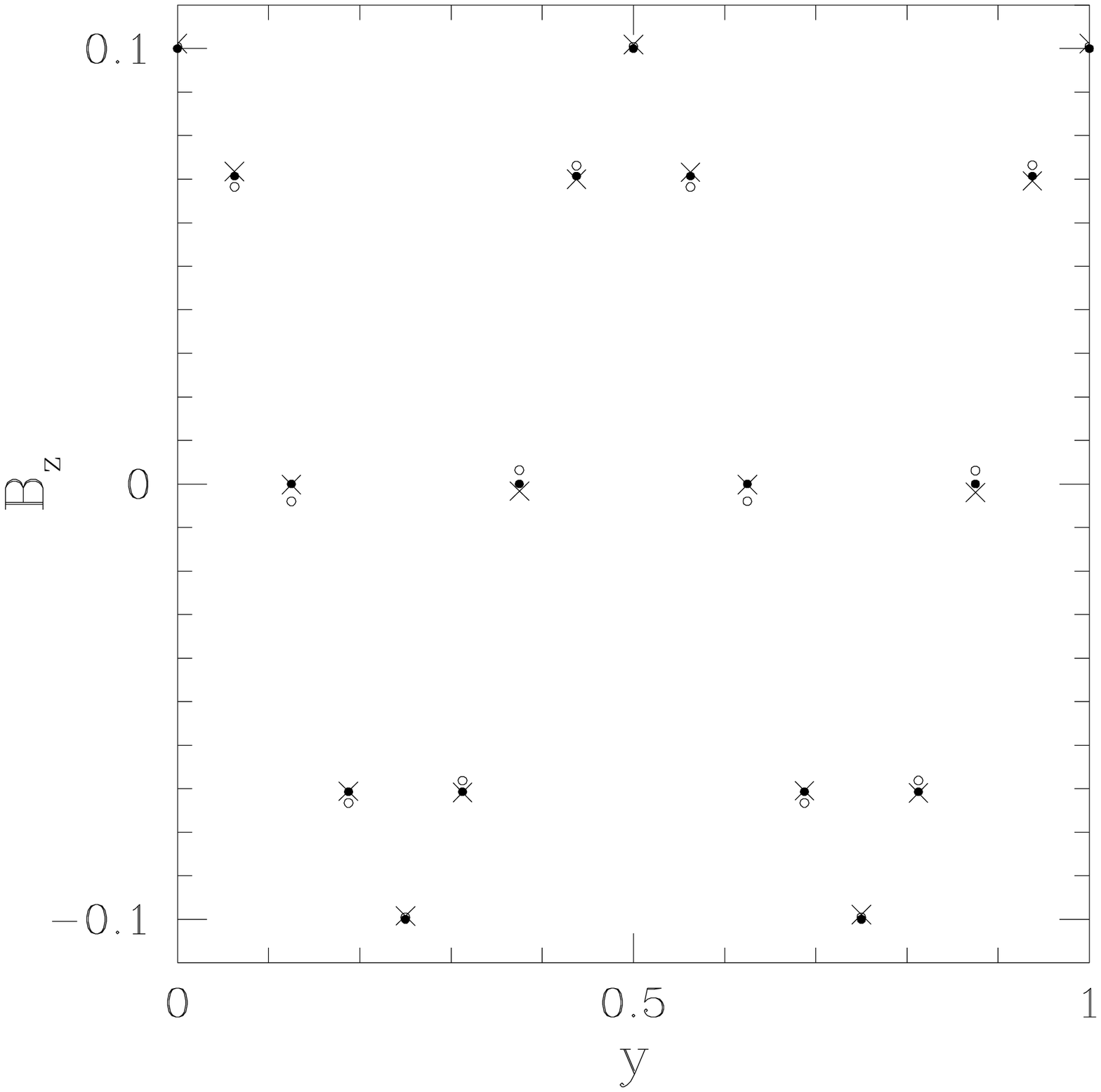} 
\caption{ \label{figalfvenb} The oblique Alfv\'en wave test from
\S~\ref{sectionalfven}.
We plot the $B_z$ field as a function of $y,$
with $x$ and $z$ equal to zero.  We show initial conditions {\em
(solid points)}, and results from simulations with polynomial finite
differences {\em (open circles)} and tuned finite differences {\em
(crosses)}.  In each simulation, the grid size is $16^3$ and the
wave has traveled 16 times around
the periodic box. The tuned finite differences yield a more precise
result.}
\end{figure}

We ran an Alfv\'en wave test where the propagation axis is oblique
to the grid axes, with the initial conditions in
Gardiner \& Stone (2005):
\begin{equation}
\V = \{ 0, \,\,\, 0.1 \sin(2 \pi x + 4 \pi y), \,\,\,
0.1 \cos(2 \pi x + 4 \pi y)\}
\end{equation}
\begin{equation}
\B = \{ 1, \,\,\, 0.1 \sin(2 \pi x + 4 \pi y), \,\,\,
0.1 \cos(2 \pi x + 4 \pi y)\}
\end{equation}
The simulation volume is a unit cube, modeled on a grid of
size $16^3$. 
The velocity field is quasi-incompressible, with the divergence removed
spectrally every 4 timesteps.
The kinetic and magnetic diffusivities
are all set to zero for this linear problem.
We ran two simulations: one with third-order polynomial finite differences
and another with third-order tuned finite differences from
Table~(\ref{tablek}).
After the wave has traveled 16 times around the periodic box, the waveform
remains almost indistinguishable from the initial conditions, with the tuned
finite differences yielding a more precise result than the polynomial
finite differences (figure~\ref{figalfvenb}).  

\section{Summary}

We have developed a new version of the constrained transport algorithm that
uses
volume-centered fields and hyperresistivity on a high-order finite difference
stencil, with tuned finite difference coefficients to enhance high-wavenumber
resolution. High-order interpolation allows implementation of staggered
dealiasing. 
Together, these measures yield a wavenumber resolution that
approaches the ideal value achieved by the spectral algorithm.

Volume centered fields are desirable because then $\V$, $\B$ and $\E$
all reside at the same grid location, allowing $\E$ to be constructed
directly from the cross product of $\V$ and $\B$ without
interpolation. For staggered fields, $\V$ and $\B$ reside at the zone
faces and $\E$ on the edges, and so constructing $\E$ involves
spatial interpolation, which reduces wavenumber resolution.

High-order stencils and tuned finite difference coefficients both enhance the
wavenumber resolution of finite differences. For a radius-three stencil with
tuned coefficients, derivatives can be computed to a relative precision of $0.5
\%$ up to a Nyquist-scaled wavenumber of $k=0.5$. Without tuning, this would be
$k= 0.34$ for a radius-three stencil. A radius-one stencil derivative such
as is used in Zeus (Stone \& Norman 1992a)
is only accurate up to $k=0.12$. The spectral derivative
is precise up to $k=1.00$, although in practice it is limited to $k=0.94$
because
of aliasing. Aliasing limits a finite-difference code to
$k=0.66$ unless the finite-difference grid shift aliasing correction is used
(\S~\ref{dealiasing}).

Hyperresistivity is desirable because it is more effective than Laplacian
resistivity in diffusing high-wavenumber modes while at the same time
preserving low-wavenumber modes. The fact that hyperresistivity can be written
as a curl allows its inclusion into CT. If Laplacian diffusivity were used
instead, too much high-wavenumber structure would be diffused for the
high-order or tuned derivatives to matter.

The resolution of the algorithm described here approaches that of a
spectral code, but because it uses finite differences, it runs faster
than a spectral code and isn't restricted to periodic boundary
conditions.  Also, since the finite differences are local, it is
easily scalable to thousands of processors. The spectral algorithm is
more difficult to scale to large numbers of processors because it involves
all-to-all communications between processors. A finite difference code
only passes information between processors whose subgrids are adjacent in
physical space. Lastly, because the code works with the magnetic field
rather than the vector potential, boundary conditions are often easier
to implement.

\begin{acknowledgements}
We received support for this work from NSF Career grant
AST99-85392, NSF grants AST03-07854 and AST06-12724, and NASA grant
NAG5-10103.  We acknowledge stimulating discussions with E. Blackman,
A. Brandenburg, B. Chandran, and J. Stone, and we also acknowledge the
referee, Wolfgang Dobler, for thorough comments that improved the paper."
\end{acknowledgements}

\appendix

\section{Appendix} \label{ctapp}

Constrained transport expresses the magnetic induction equation as a pure curl
plus a divergence diffusivity: $\partial_t \B = \bnabla \times {\bf F} +
\eta_{\smf D} \bnabla \bnabla \cdot \B,$ where ${\bf F}$ is defined in
equation~(\ref{eqctb}).  The $\eta_{\smf D}$ term serves to diffuse away
magnetic divergence, and the finite differences are arranged so that $\bnabla
\cdot \bnabla \times {\bf F} = 0.$ Thus, the curl term does not contribute to
the evolution of the magnetic divergence, and if the initial conditions are
divergence-free, the magnetic divergence remains zero throughout the evolution.

\begin{figure}[!] \plotone{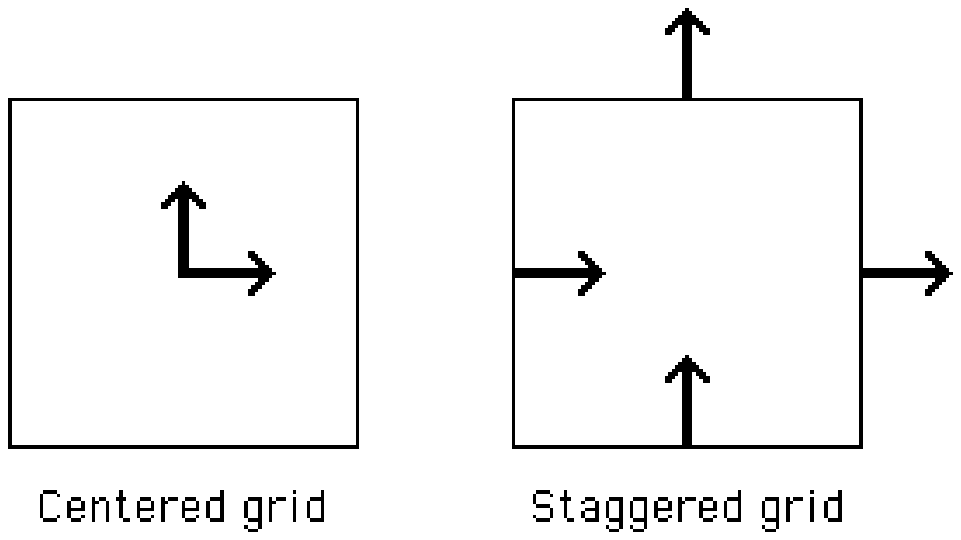}
\caption{
\label{figstagger}
We show the configuration for zone-centered and zone-staggered fields
in two dimensions, where the generalization to three dimensions is
straightforward. {\em (Left)} A zone-centered vector field, where both the
$\hat{X}$ and $\hat{Y}$ vectors reside at zone centers. {\em (Right)} A
zone-staggered vector field, where $\hat{X}$ vectors reside on
$\hat{X}$ faces and $\hat{Y}$ vectors reside on $\hat{Y}$ faces.}
\end{figure}

To see how constrained transport works,
denote the vector field by
${\bf F_{\smf i,j,k}}$ = $\{F^x_{\smf i,j,k}, F^y_{\smf i,j,k},
F^z_{i,j,k}\},$ where $\{i,j,k\}$ are integers specifying the locations of
grid cell centers. There are two basic grid types: ``centered" and
``staggered" (figure~\ref{figstagger}).
For a centered grid, scalar and vector quantities are
located at cell centers. For a staggered grid,
scalar quantities are located at
cell centers and vector quantities at cell faces.
For instance, we would index the components of ${\bf F}$ as
$\{F^x_{\smf i+1/2,j,k}, F^y_{\smf i,j+1/2,k}, F^z_{\smf i,j,k+1/2}\}.$

The finite divergence divergence of the curl of ${\bf F}$ is $\bnabla \cdot
\bnabla \times {\bf F} = \epsilon_{ijk} \partial_i \partial_j {\bf F}_k,$ which
consists of terms such as $(\partial_1 \partial_2 - \partial_2 \partial_1)
{\bf F}_k.$ One can straightforwardly see that this is zero for finite
differences of the form eq.~\ref{stencil} for both
centered and staggered grids. Thus, constrained
transport can be coordinated with high-order and tuned finite differences,
as well as with hyperresistivity.

For a staggered grid, the ``${\bf F}$" vectors are located at cell
edges, whereas the $\V$ and $\B$ vectors from which they are constructed
are found at cell faces. A staggered grid CT scheme therefore involves
spatial interpolation, one example
being the method of characteristics scheme for
time-interpolating Alfv\'en waves. We use volume-centered fields and
Runge-Kutta timestepping because,
among other reasons, no interpolation is required.

\end{document}